\documentclass[11pt]{article}
\usepackage{moriond,epsfig}
\usepackage{epstopdf}
\usepackage{amssymb}

\def\be{\begin{equation}}
\def\ee{\end{equation}}
\def\bea{\begin{eqnarray}}
\def\eea{\end{eqnarray}}

\begin{document}
\vspace*{4cm}
\title{LARGE-ANGLE ANOMALIES IN THE MICROWAVE BACKGROUND}

\author{ E.F. BUNN }

\address{Physics Department, University of Richmond\\
28 Westhampton Way, Richmond, VA  23173, USA}

\maketitle\abstracts{
Several claims have been made of anomalies in the large-angle properties of the cosmic microwave
background anisotropy as measured by WMAP.  In most cases, the statistical significance of these
anomalies is hard or even impossible to assess, due to the fact that the statistics used to quantify
the anomalies were chosen a posteriori.  On the other hand, the possibility of detecting new physics on the largest observable scales is so exciting that, in my opinion, it is worthwhile to examine the claims carefully.  I will focus on three particular claims: the lack of large-angle power, the north-south power asymmetry, and multipole alignments.  In all cases, the problem of a posteriori statistics can best be solved by finding a new data set that probes similar physical scales to the large-angle CMB.  This is a difficult task, but there are some possible routes to achieving it. }

\section{Introduction}
Our understanding of cosmology has advanced extremely rapidly
in the past decade, due in large part to
observations of cosmic microwave background (CMB) anisotropy,
particularly the data from the Wilkinson Microwave Anisotropy
Probe (WMAP).\cite{wmap1yr1,wmap1yr2,wmap1,wmap7yrbasic}
As a result of these and other observations, a
``standard model'' of cosmology has emerged, consisting of a Universe dominated
by dark energy and cold dark matter, with a nearly scale-invariant
spectrum of Gaussian adiabatic perturbations\cite{wmap7yrparams,wmap7yrinterp} of the sort
that would naturally be produced in an inflationary epoch.  

The all-sky CMB maps made by WMAP provide a unique window on the Universe, probing larger scales and earlier times than any other data.  Not surprisingly, therefore, researchers have scrutinized the maps carefully for evidence of nonstandard phenomena on large scales.   In most ways, there is remarkable agreement between the standard model and the large-angle properties of the WMAP data; however, 
several anomalies have been noted on the largest
 scales, including among others a lack of large-scale power,\cite{wmap1yr2,copi2,dOCTZH} alignment of low-order multipoles,\cite{dOCTZH,schwarz,copi1,hajian} and hemispheric asymmetries\cite{eriksen2004,freeman,hansen}.  The WMAP collaboration has conducted a thorough review and analysis of the claimed anomalies.\cite{wmap7yranomalies}

The significance of and explanations for these puzzles
are uncertain, largely because of the problem of a posteriori statistics.  The typical sequence of events in the discovery of a CMB anomaly is as follows: some unusual feature is noticed in the data, and afterwards (a posteriori) a statistic is devised to quantify the unusualness of this feature.  The $p$-values from such a statistic cannot be taken at face value: in any moderately large data set, it is always possible to find something that looks odd, and a statistic engineered to capture that oddness will have an artificially low probability.  Chance fluctuations can therefore incorrectly seem to be in need of explanation.

\begin{figure}[t]
\centerline{\raisebox{1cm}{\epsfig{width=7.5cm,figure=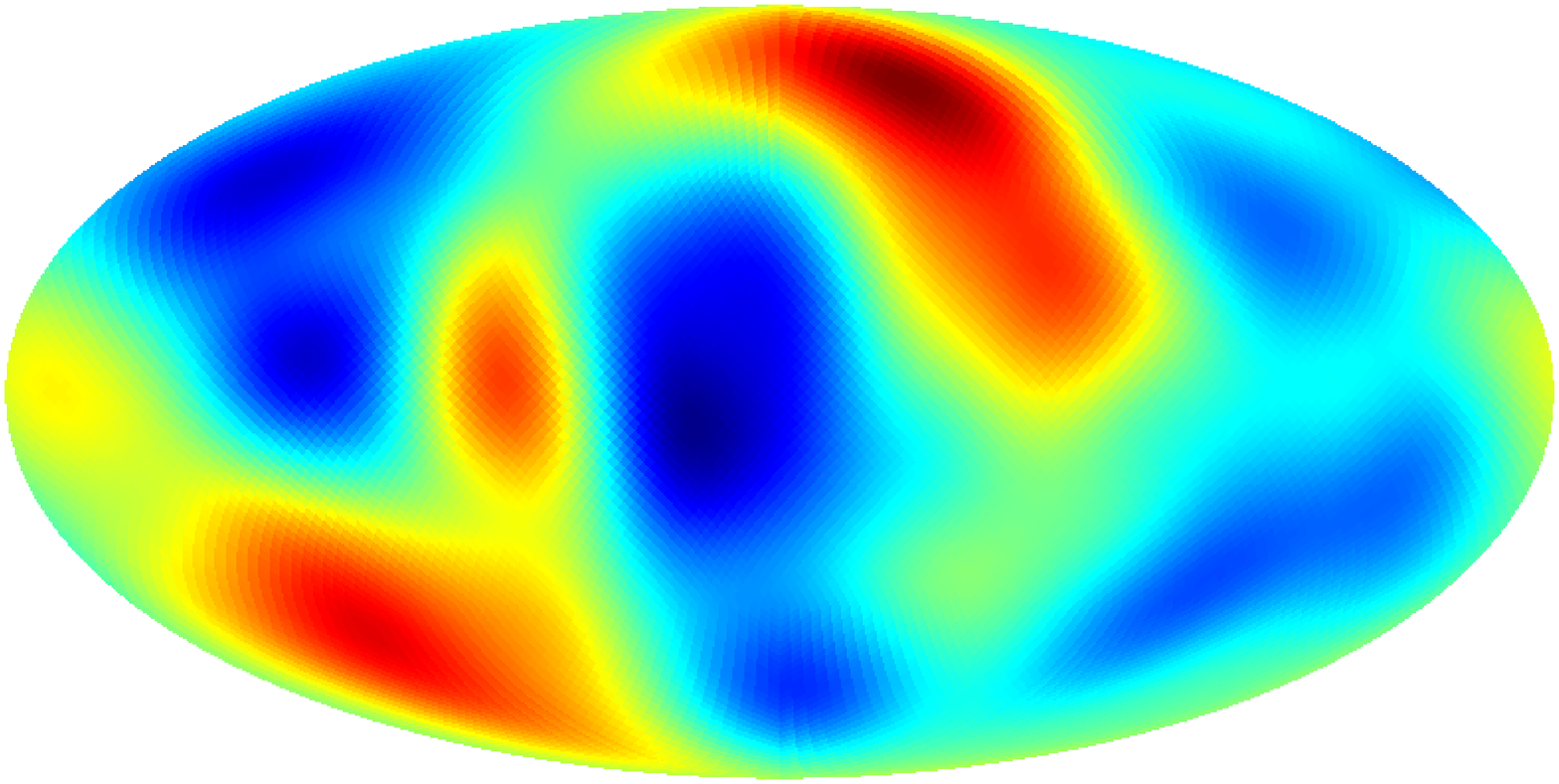}}
\epsfig{width=7.5cm,figure=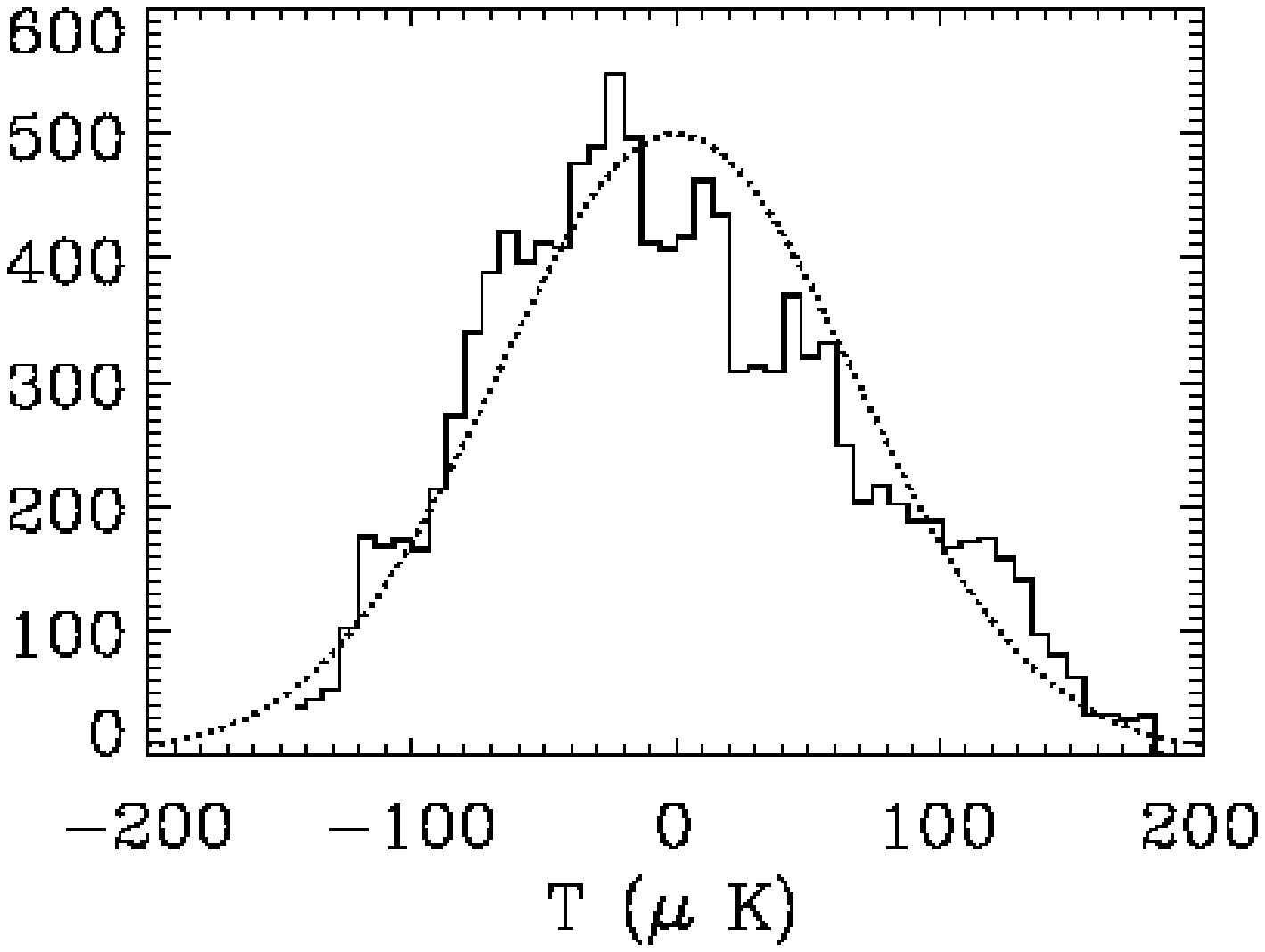}}

\caption{A toy example of the dangers of a posteriori statistics.  The left panel shows a Gaussian, statistically isotropic simulation of a CMB map, smoothed to show only large-scale anisotropy.  As the right-panel shows, the one-point probability distribution of the anisotropy has an unusually high skewness.  This skewness can be ``explained'' by noting the presence of a pair of almost perfectly antipodal extreme hot spots.}
\label{fig:toy}
\end{figure}

Figure \ref{fig:toy} illustrates this with a simple toy example.  A Gaussian random map was made that by chance has noticeable positive skewness.  Someone looking for an explanation for this skewness might be tempted to examine the most extreme hot spots in the map and would find that they are almost perfectly antipodal.  The probability that a map's two most extreme hot spots are as far apart as in this map is less than 1\%.  One might be tempted to speculate on the possible explanations for this unlikely pair of hot spots, but of course there is none. 

One can (and from a formal statistical point of view, arguably
one must) dismiss the entire subject of CMB anomalies because of this problem, but I believe that a more nuanced view is called for.  Scientists often by necessity use non-rigorous (or even ``invalid'') statistical methods, especially in preliminary analyses.  As long as we maintain a skeptical stance and seek further tests that can be done of any hypotheses that result from such an analysis, these methods can yield fruitful insights.  Considering the importance of finding ways to test the largest-scale properties of the Universe, I suggest that is neither necessary nor wise to dismiss the subject out of hand.

In this paper, I will not discuss all of the claims of anomalies but rather focus on the three that in various ways seem to me most interesting: the large-scale power deficit, evidence for hemispheric power modulation, and alignment of low-order multipoles.  I will also discuss possible future directions for testing hypotheses arising from these anomalies.

\section{Lack of large-scale power}

\begin{figure}[t]
\centerline{\epsfig{width=7.5cm,figure=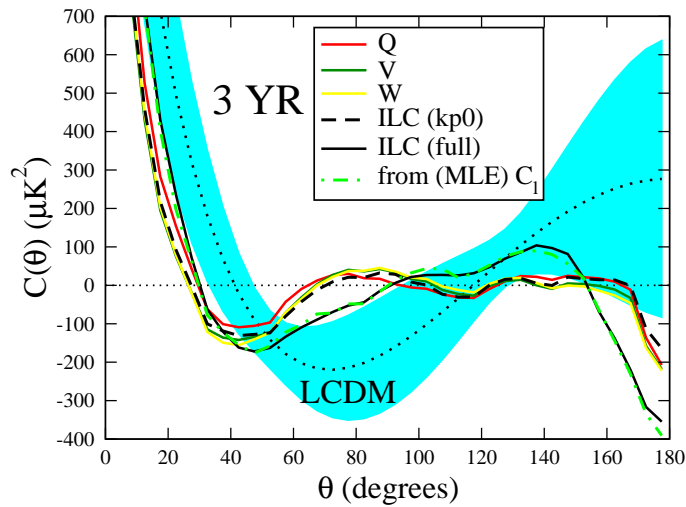,angle=270}}
\caption{From Copi et al.\protect\cite{copi2} two-point correlation function for the three-year WMAP data.  The blue band shows the expected range from simulations.}
\label{fig:corrfunc}
\end{figure}

Ever since the first all-sky CMB maps were made by the COBE satellite,\cite{smoot92} questions have been raised about the low amplitude of fluctuations on the largest angular scales.  In the observed angular power spectrum, the quadrupole (the largest-scale data point) is lower than expected, although given the large cosmic variance in the quadrupole as well as the need to mask part of the sky, this discrepancy is not extremely statistically significant.    The lack of large-scale power appears more striking when viewed in real space rather than the spherical harmonic space of the power spectrum.  As Figure \ref{fig:corrfunc} illustrates, the two-point correlation function is very close to zero for all angles $\theta\gtrsim 60^\circ$, unlike typical simulations.

To quantify this behavior, the statistic $S_{1/2}$ is defined to be\cite{wmap1yrparams}
\begin{equation}
S_{1/2}=\int_{-1}^{1/2}[C(\theta)]^2\,d\cos\theta.
\end{equation}
The value of this statistic is low compared to simulations at a confidence level of approximately 99.8\%.\cite{copi2}

This statistic is of course a prime example of an ``invalid'' a posteriori statistic, chosen to quantify an already-noticed odd feature in the data, so this confidence level must be interpreted with skepticism.  On the other hand, the two-point correlation function is one of the simplest and most natural quantities to compute from a CMB sky map --- for many years, in fact, it was \textit{the} chief way CMB anisotropy was quantified.  The qualitative difference between the observed WMAP correlation function and theoretical predictions is therefore intriguing.

\begin{figure}[t]
\centerline{\epsfig{width=7.5cm,figure=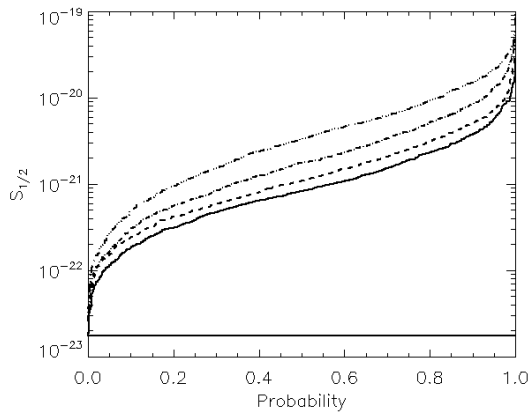}}
\caption{From Bunn \& Bourdon\protect\cite{bunnbourdon}.  The solid curve shows the cumulative probability distribution for the statistic $S_{1/2}$.  The dashed curves show the distributions for various anisotropic (ellipsoidal) models.  At any given value of $S_{1/2}$, the probability is always largest for the standard model.}
\label{fig:shalfprob}
\end{figure}

If we tentatively assume that the lack of large-scale power does require an explanation, then it is natural to ask what form that explanation might take.  We can rule out one broad class of explanations, namely those that involve a statistically independent additive contaminant to the data.\cite{copi2,bunnbourdon}  The reason is that such a contaminant always biases the expected amount of large-scale power up, rendering the low observed value less likely, not more. It is clear that a contaminant always adds power in an rms sense: the quadrupole (or any other multipole) is simply the quadrature sum of the contributions from the true CMB anisotropy and the contaminant.  But I am making the stronger statement that the entire probability distribution shifts in such a way that a low value of the power becomes more improbable (e.g., Figure \ref{fig:shalfprob}).  This statement is true whether the lack of large-scale power is quantified by the observed quadrupole\cite{copi2} or by the statistic $S_{1/2}$.\cite{bunnbourdon}  It is independent of any assumptions about the statistics of the contaminant, as long as it is independent of the intrinsic CMB anisotropy.

For example, this result rules out an undiagnosed foreground as an explanation for the power deficit.  While it is possible that a foreground could cancel the intrinsic large-scale power, such a chance cancellation is always \textit{less} likely than the power coming out low without the contaminant. In fact, if a foreground contaminant is invoked for some other reason (e.g., to explain one of the other anomalies), it will exacerbate the large-scale power deficit problem.

 In addition to foreground contaminants, some more exotic models also fall into the category ruled out by this model, such as ellipsoidal models\cite{ellipsoid}, some models with large-scale magnetic fields\cite{barrow}, etc.

\section{Hemisphere asymmetry}

\begin{figure}[t]
\centerline{\epsfig{height=7.5cm,figure=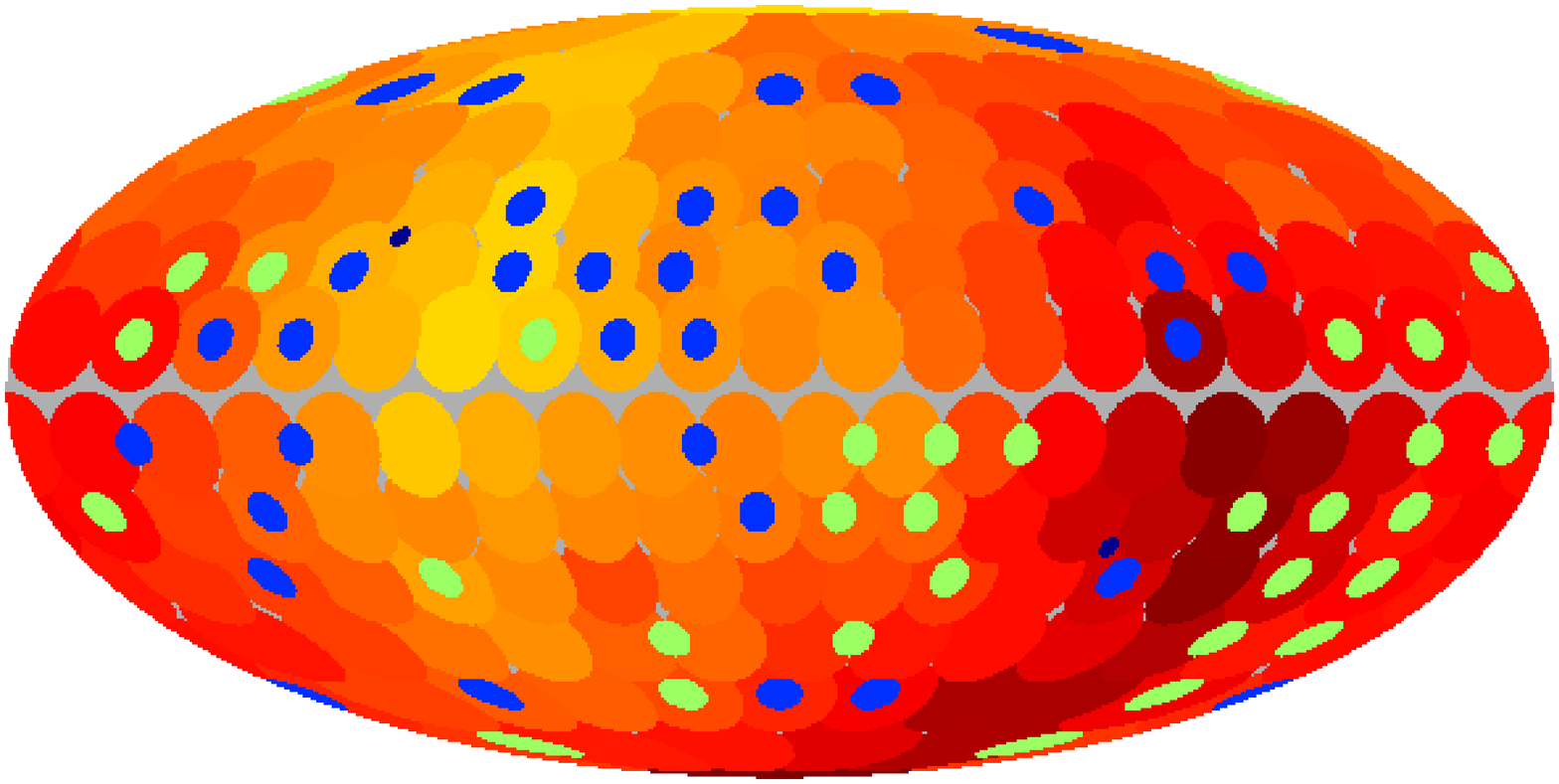,angle=90}
\epsfig{height=7.5cm,figure=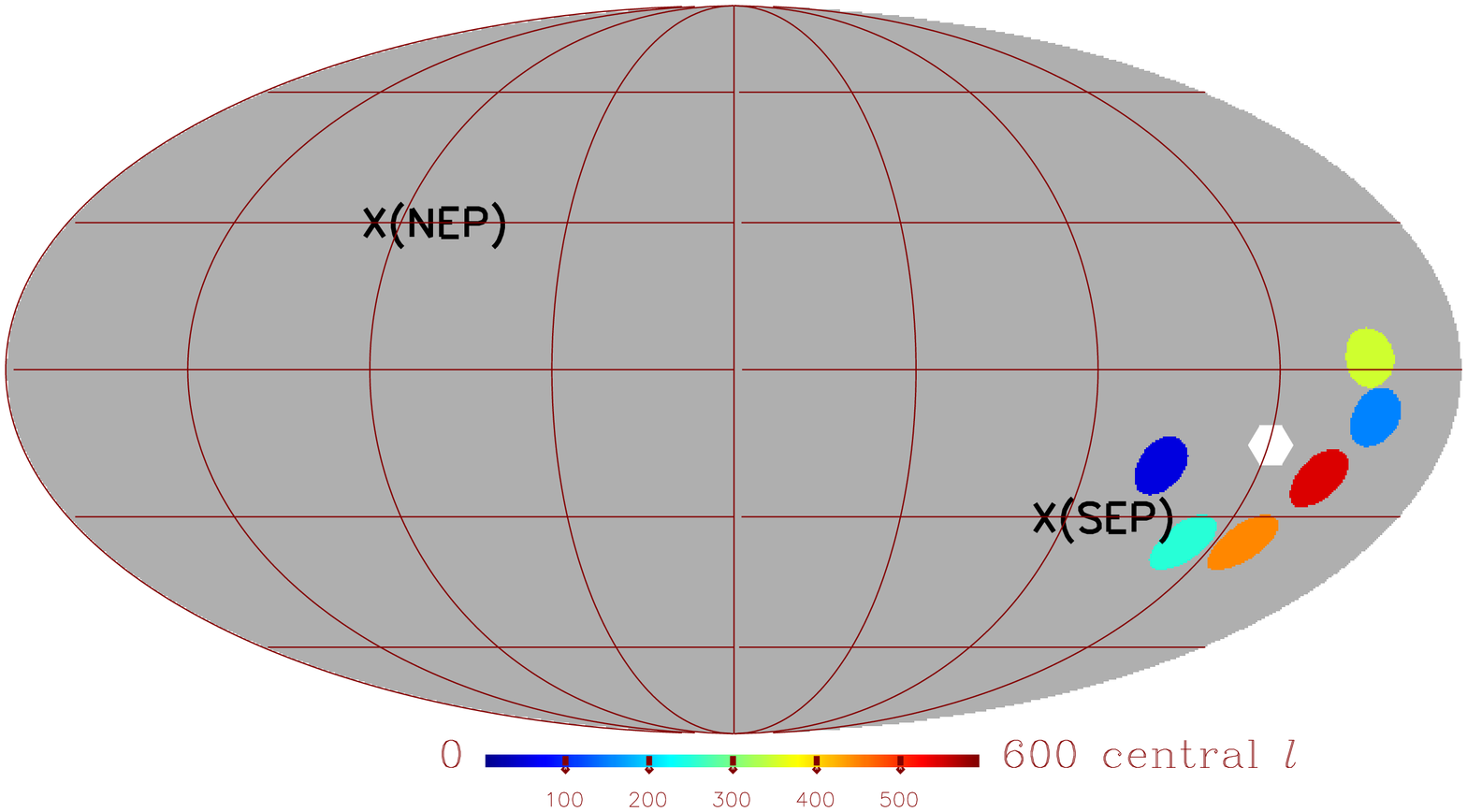,angle=90}}
\caption{The left panel (from Eriksen et al.\protect\cite{eriksen2004}) illustrates the hemisphere 
asymmetry seen in WMAP.  The color of eac large disk indicates the ratio of power in a hemisphere centered on the disk to power in the opposite hemisphere, considering only multipoles $l=2$-$63$.  
The right panel (from Hansen et al.\protect\cite{hansen}) shows the directions yielding maximum power asymmetry for different ranges of multipoles.}
\label{fig:hemisphere}
\end{figure}

In the standard cosmological paradigm, CMB anisotropy is generated by a statistically isotropic random process, meaning that all directions should be, on average, identical.  However, there appears to be a large-scale modulation in power in the WMAP sky maps: more fluctuation power is seen in one hemisphere than in the opposite hemisphere.  Figure \ref{fig:hemisphere}a illustrates this power asymmetry.

Figure \ref{fig:hemisphere}b shows an important test of this anomaly performed by Hansen et al.\cite{hansen}.  The WMAP data are filtered into non-overlapping ranges of multipoles: $l=2-101, 102-201,\ldots$.  In each case, the direction is found that maximizes the power asymmetry (i.e., maximizing the ratio of power in a hemisphere centered on that direction to power in the opposite hemisphere).  In the standard model, these directions should all be independent random variables, but they are clearly closely correlated with each other.  Even if the initial detection of a power asymmetry is contaminated by the problem of a posteriori statistics, this close correlation provides a largely independent test that is relatively free of this contamination.

I believe that this test provides strong evidence that the power asymmetry is truly present in the data, but this does not mean that it is cosmological.  As indicated in the Figure, the power-maximizing direction is quite close to the south ecliptic pole.  If this alignment is not a coincidence, then the hemisphere asymmetry has a local cause, perhaps related to the WMAP scan strategy.

\section{Multipole alignments}

\begin{figure}[t]
\centerline{\epsfig{width=7.5cm,figure=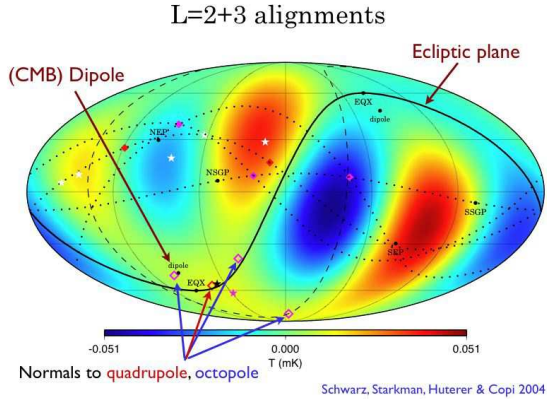}}
\caption{From Schwarz et al.\protect\cite{schwarz}  The quadrupole plus octupole of the WMAP data.  Several directions that can be computed from these multipoles are indicated, along with the orientations of the ecliptic and dipole.}
\label{fig:alignment}
\end{figure}

The alignment of low-order multipoles, especially the quadrupole and octopole ($l=2,3$) may have received the most attention of all of the claimed anomalies.  For each multipole, one can define a plane in which the fluctuations preferentially lie, using an angular-momentum statistic\cite{dOCTZH} or multipole vectors\cite{copi1}.  These directions are expected to be independent of each other but are surprisingly closely aligned.  In addition, the directions perpendicular to these planes are close to both the CMB dipole and the ecliptic plane (see Figure \ref{fig:alignment}).  Depending on which of these surprises one chooses to consider and how one chooses to quantify them, it is easy to get $p$-values of $10^{-3}$ or less\cite{schwarz}.  (Of course, as with the hemisphere asymmetry, if there is a cosmological explanation for the alignments, then the alignment with the ecliptic must be a mere coincidence.)

Yet again the problem of a posteriori statistics rears its head.  The lowest $p$-values arise from considering alignment of things that have been seen to be aligned.  It is difficult to know how to correct this for the various alignments that could have been seen but weren't.  Reasonable people can (and do)  differ over how much weight to give to the various multipole alignments.  In my opinion, it is impossible to be confident that the observed alignments are significant, but it is reasonable to use them to generate hypotheses for future examination and then look for ways to test these hypotheses with new data sets that are independent of the large-angle CMB.

\section{Possible explanations and future tests}

If we tentatively assume that a given anomaly is ``real'' (i.e., not merely a statistical fluctuation amplified by an a posteriori statistic), then it is natural to ask what  explanations might be possible.  Possibilities include systematic errors, foreground contaminants (although not in the case of the power deficit), or more exotic explanations involving new physics.  Examples from the latter category are theories that define a preferred direction in space, either through spontaneous isotropy breaking\cite{gordon} or by the presence of a vector field during inflation\cite{ackerman}.  

\begin{figure}[t]
\begin{minipage}[b]{0.45\linewidth}
\centerline{\epsfig{width=6cm,figure=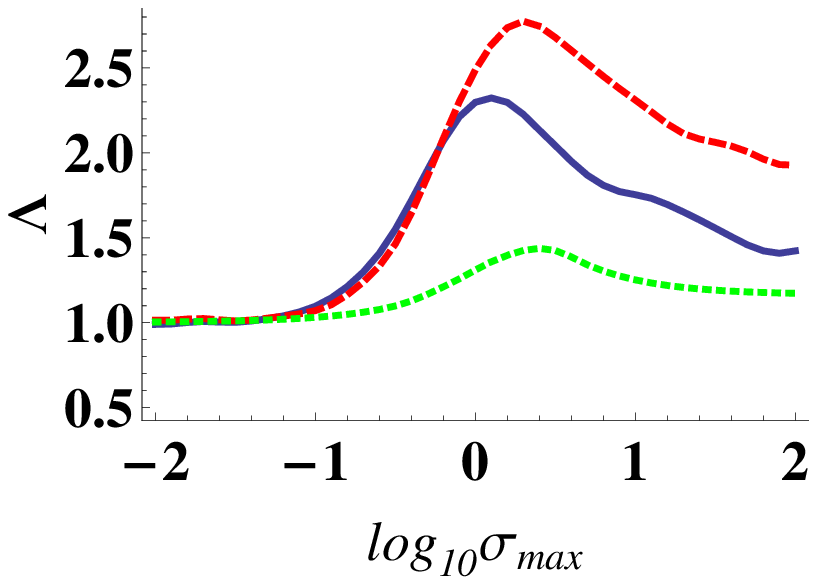}}
\caption{An illustration of the use of Bayesian evidence to decide whether anomalies in the data warrant adoption of a more complicated model.  The quantity plotted is the Bayesian evidence ratio comparing various anisotropic models (two based on the idea of spontaneous isotropy breaking and one involving a preferred direction during inflation) with the standard model. The statistic used in computing the evidence ratio is based on the multipole vector method of quantifying quadrupole-octopole alignment. For further details, see Zheng \& Bunn.\protect\cite{zhengbunn}}
\label{fig:bayes}
\end{minipage}
\hspace{0.1\linewidth}
\begin{minipage}[b]{0.45\linewidth}
\centerline{\epsfig{width=6cm,figure=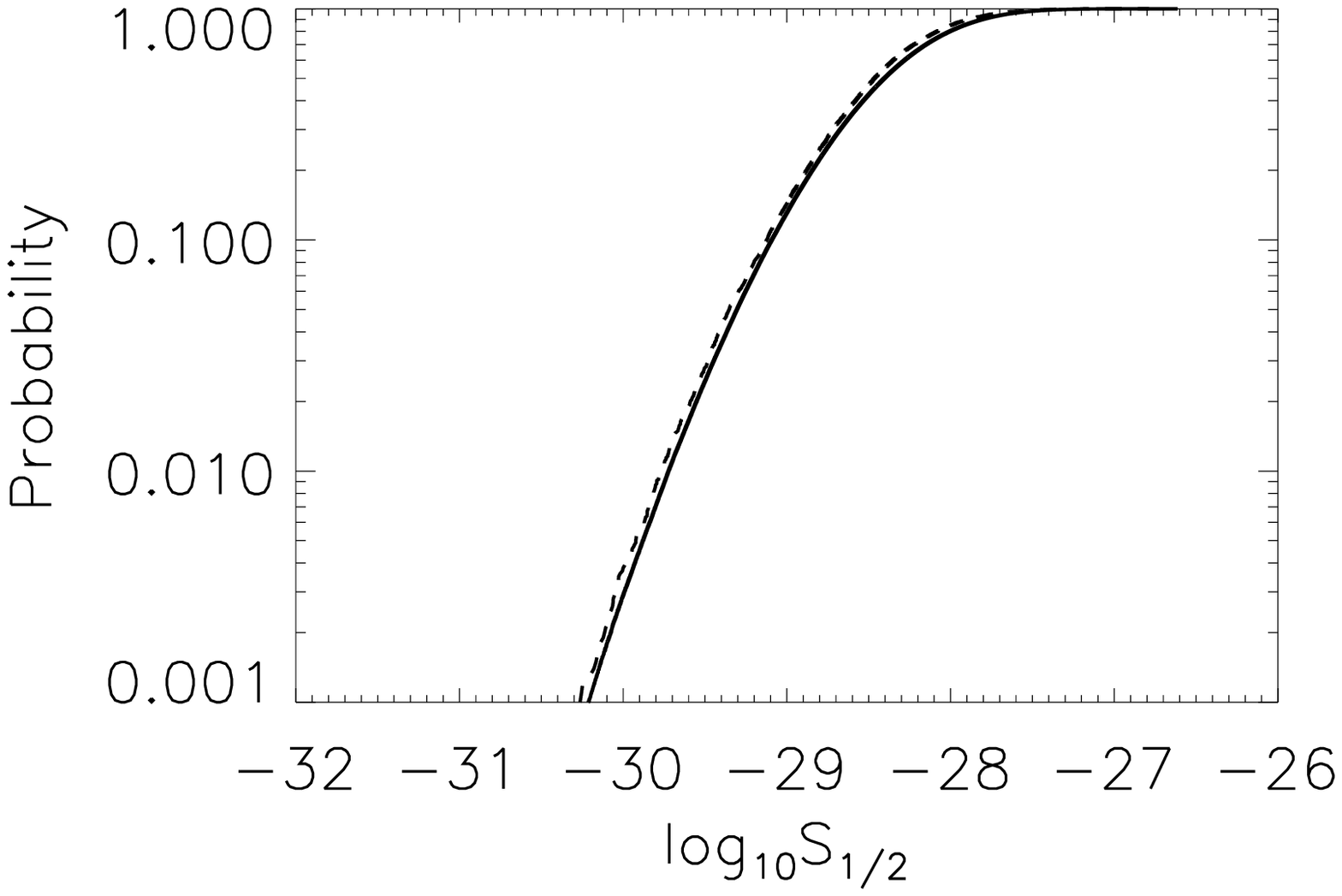}}
\caption{The cumulative probability distribution of the statistic $S_{1/2}$ for E-mode polarization, as predicted by the standard model.  The solid curve shows the overall probability distribution, derived from simulations based on a standard $\Lambda$CDM model.  The dashed curve is the probability distribution conditioned on extremely low values (first percentile) of the {\it temperature} $S_{1/2}$ statistic.  The close correspondence of the curves shows that a low value of $S_{1/2}$ for polarization may be regarded as independent, to a good approximation, of the measurement of a low $S_{1/2}$ in temperature.}
\label{fig:shalfpol}
\end{minipage}
\end{figure}

Because the alternative theories generally have additional free parameters (and usually include the standard model as a limiting case), they typically can provide better fits to the data.  A model selection criterion is required to decide whether the improved goodness of fit is worth the ``cost'' of a more complex theory.  Perhaps the most natural such criterion is the Bayesian evidence ratio\cite{jeffreys,landmag,eriksen2007,hoftuft,liddle,efstathioubayes,zhengbunn}, which is essentially the factor by which the posterior probability ratio of two theories is increased, in comparison to the prior ratio, by the acquisition of the new data.  The evidence ratio automatically disfavors complicated theories (i.e., those with large parameter spaces) unless the improvement in fit is correspondingly large; in other words, it automatically incorporates a form of Occam's razor.  Recent work\cite{hoftuft,zhengbunn} has attempted to quantify the Bayesian evidence ratios for certain classes of theories.  Figure \ref{fig:bayes} illustrates an example, in which we quantify the degree to which the quadrupole-octopole alignment improves the likelihood of spontaneous isotropy breaking and preferred direction models.  Although the evidence ratios exceed 1, indicating that the more complicated models go up in probability as a result of the multipole alignment, the improvement is extremely modest.  In general, one does not pay much attention to Bayesian evidence ratios unless they are far larger than these values\cite{jeffreys}.

Because of the uncertainties surrounding the interpretation of the statistics of the various CMB anomalies, they should be regarded chiefly as potentially useful guides in formulating hypotheses for further testing.  The essential next step, therefore, is to find new data sets that can be used to test any such hypotheses.  To be specific, we need to find data sets that probe comparable physical scales to the large-angle CMB (i.e., gigaparsec scales in comoving coordinates) but that are independent of the CMB anisotropy modes, which have already been measured to the cosmic-variance limit. Finding such data sets is nontrivial, of course, but it may not be impossible.

The first natural place to look is to CMB polarization.  For any given anomaly, one can imagine devising (a priori) statistical tests to be performed on polarization maps to look for the anomaly's presence.  For example, one can compute the two-point correlation function of a map of E-type CMB polarization, and see if it shows the same lack of large-angle power as the temperature anisotropy.  Because there are correlations between temperature and E polarization, this is not, strictly speaking, an independent test, but in practice it is nearly so: as Figure \ref{fig:shalfpol} shows, the predicted probability distribution for an $S_{1/2}$ statistic computed from a polarization map is essentially independent of the value of $S_{1/2}$ for temperature.  When CMB polarization data are good enough to allow reliable estimation of the correlation function, we can compute $S_{1/2}$.  If it is anomalously low, then we have found independent evidence that this puzzle requires an explanation.

Unfortunately, this test is likely to be of less value than it might initially appear: the bulk of the large-angle power in CMB polarization data comes from photons that last scattered at low redshift (after reionization), and hence probes far smaller length scales than the corresponding temperature data.  Thus even if there is a cosmological explanation for the lack of large-scale temperature correlations, we would probably not expect to find confirmation of it in polarization. The same conclusion would likely apply to tests of other anomalies.   A polarization map does in principle contain information on large length scales that is independent of the temperature data, but it is not obvious (at least to me) that this information can be separated from the reionization signal in a way that would allow a clear test of the anomalies.  

There are other possibilities for independent probes of perturbations on gigaparsec scales.  Since the CMB anisotropy primarily probes the surface of last scattering, methods that sample the interior of our horizon volume will generically provide independent data sets.  One method that might prove promising in the future is the Kamionkowski-Loeb effect:\cite{kamloeb} by measuring the polarization of the Sunyaev-Zel'dovich-scattered photons coming from a galaxy cluster, one can infer the CMB temperature quadrupole measured at that cluster's location and look-back time.  A sample of such cluster measurements can be used to reconstruct modes on length scales corresponding to the CMB modes at $l\sim 5$.\cite{bunnrq,abramorq}  This is a challenging task, but the generation of telescopes currently being developed, such as SPTPol and ACTPol, are capable of achieving it.

\section{Conclusions}

The subject of large-angle CMB anomalies remains controversial, largely because of the difficulty in interpreting a posteriori statistics.  While reasonable people can and do conclude that the correct attitude is to dismiss the subject entirely, I believe that a more nuanced view is appropriate, in which we view some anomalies as providing hints of possible new directions to explore.  With this attitude, it is of course essential to seek rigorous tests of any hypotheses generated.  Such tests may be difficult but not impossible to find.

\section*{Acknowledgments}
This work was supported by US National Science Foundation Awards 0507395 and 0908319.  I thank the Laboratoire Astroparticule-Cosmologie of the Universit\'e Paris VII for their hospitality during some of the time
this work was prepared.

\section*{References}
\bibliography{bunn-moriond-2010}
%\begin{thebibliography}{99}
%\bibitem{ja}C Jarlskog in {\em CP Violation}, ed. C Jarlskog
%(World Scientific, Singapore, 1988).

%\bibitem{ma}L. Maiani, \Journal{\PLB}{62}{183}{1976}.

%\bibitem{bu}J.D. Bjorken and I. Dunietz, \Journal{\PRD}{36}{2109}{1987}.

%\bibitem{bd}C.D. Buchanan {\it et al}, \Journal{\PRD}{45}{4088}{1992}.

%\end{thebibliography}

\end{document}